\documentclass[prb,twocolumn,showpacs,superscriptaddress,preprintnumbers,amssymb]{revtex4}
\usepackage{graphicx}
\usepackage{dcolumn}
\usepackage{bm}
\usepackage{wasysym}

\newcommand{\s}{{\sigma}}

\newcommand{\beq}{\begin{equation}}
\newcommand{\eeq}{\end{equation}}
\newcommand{\beqn}{\begin{eqnarray}}
\newcommand{\eeqn}{\end{eqnarray}}

\begin{document}
\title{Dimensional reduction in superconducting arrays and frustrated magnets}
\author{Cenke Xu}
\affiliation{Department of Physics, University of California,
Berkeley, CA 94720}
\author{J.~E.~Moore}
\affiliation{Department of Physics, University of California,
Berkeley, CA 94720} \affiliation{Materials Sciences Division,
Lawrence Berkeley National Laboratory, Berkeley, CA 94720}
\date{\today}
\begin{abstract}
Some frustrated magnets and superconducting arrays possess unusual symmetries that cause the free energy or other physics of a $D$-dimensional quantum or classical problem to be that of a different problem in a reduced dimension $d<D$.  Examples in two spatial dimensions include the square-lattice $p+ip$ superconducting array, the Heisenberg antiferromagnet on the checkerboard lattice (studied by a combination of $1/S$ expansion and numerical transfer matrix), and the ring-exchange superconducting array.   Physical consequences are discussed both for ``weak'' dimensional reduction, which appears only in the ground state degeneracy, and ``strong'' dimensional reduction, which applies throughout the phase diagram.  The ``strong'' dimensional reduction cases have the full lattice symmetry and do not decouple into independent chains, but their phase diagrams, self-dualities, and correlation functions indicate a reduced effective dimensionality.  We find a general phase diagram for quantum dimensional reduction models in two quantum dimensions with $N$-fold anisotropy, and obtain the Kosterlitz-Thouless-like phase transition as a deconfinement of dipoles of 3D solitons.
\end{abstract}
\maketitle

\section{Introduction}

The subject of this paper is a set of problems in condensed matter physics where unusual symmetries generate three- or four-body interactions without two-body interactions, even though the underlying Coulomb interaction is a two-body interaction.  The examples that we study are drawn from models of frustrated Heisenberg magnets and complex superconducting arrays.  Our focus here is on a dramatic consequence of the four-body interaction that appears naturally in these systems: the apparent dimensionality of the system is different from its effective dimensionality, either for the system's phase transition or its ground state degeneracy, or both.  Most possible experiments on the systems we discuss should measure correlations that are sensitive to the effective dimensionality.

Some examples of this phenomenon are relatively simple and well understood.  The $1/S$ expansion approach to the 2D checkerboard Heisenberg antiferromagnet~\cite{henley, tchernyshyov} labels the infinite degenerate classical ($S\rightarrow \infty$) ground states by local Ising variables.  Including $1/S$ corrections, which come from the zero-point energy of spin-wave fluctuations, partially splits the extensive $S \rightarrow \infty$ degeneracy and generates a four-spin energy function on the Ising variables.  As reviewed in Section II, this $1/S$ energy function supports a 1D degeneracy of ground states, even though the model is two-dimensional and symmetric $x \leftrightarrow y$.  However, as we confirm numerically, this effective model has a finite-temperature phase transition (an example of ``order by disorder''~\cite{chandra}) as usual in 2D.  Only the ground state degeneracy suggests a reduced dimensionality, and we label this ``weak'' dimensional reduction.

Most of the discussion in this paper is toward understanding more subtle examples, both quantum and classical, and in arbitrary dimension, where dimensional reduction persists for all couplings.  An example appearing in a 2D quantum model for $p+ip$ superconducting arrays shows an exact self-duality identical to that of the 2D classical Ising model, plus other signatures of 2D classical or 1D quantum physics.  This model is one of the few cases where it is possible to obtain exact information on a two-dimensional quantum phase transition.

There are several physical problems (Section II) where the above physics may be tested experimentally.  The simplest experimental signature in lattice models is correlations that are highly directional: in general the long-wavelength correlations near a quantum critical point are approximately isotropic, but that is not the case for the critical points we discuss.  Although the correlations have the full symmetry of the lattice (there is no decoupling into ``chains''), they do not have the continuous rotational symmetry of ordinary critical points.

For clarity, we give a classical, fully solvable example of dimensional reduction (arising in a model of superconducting arrays), so that the basic idea and physical significance will be clear.  This model has no finite-coupling phase transition; most examples in later sections have unusual classical or quantum phase transitions.  The simplest example of what we mean by ``dimensional reduction'' is the 2D classical theory on the square lattice with an Ising spin $s_i = \pm 1$ at each site and plaquette interactions:
\beq
-\beta E = K \sum_\square s^\square_1 s^\square_2 s^\square_3 s^\square_4
\label{clas2d}
\eeq
where the sum is over elementary squares and the four spins $s_i^\square$ are the corners of square $\square$.  In Cartesian coordinates
\beq
-\beta E = K \sum_{i,j \in \mathbb{Z}} s_{i,j} s_{i+1,j} s_{i+1,j+1} s_{i,j+1}
\eeq
Section II reviews how this model gives the effective description of $p\pm ip$ superconducting arrays~\cite{moorelee}.

This model has exactly the same free energy per site as the 1D nearest-neighbor Ising model $-\beta E = K \sum_i s_i s_{i+1}$.  To see this, note that just as in the 1D Ising model all the bonds $b_i = s_i s_{i+1}$ can be specified independently (up to two spins at the boundary), in the 2D plaquette model (\ref{clas2d}) all face variables
\beq
f_\square = s^\square_1 s^\square_2 s^\square_3 s^\square_4
\eeq
can be specified independently (up to a single horizontal line and single vertical line of spins).

\begin{figure}
\includegraphics[width=1.5in]{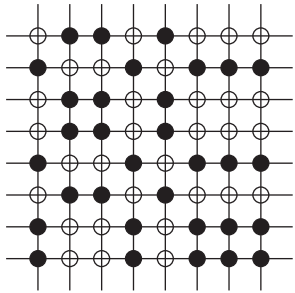}
\caption{A typical ground state configuration of Ising variables
(open or filled circles) for plaquette interactions, as appear in a $p\pm ip$ superconducting array.  Note that fixing one row and one column determines the rest of the lattice via the constraint of 0, 2, 4 up-spins per square. } \label{figgs}
\end{figure}

Given arbitrary values of the spin on one horizontal line and one vertical line, there is a unique ground state of the system compatible with these values (Fig.~\ref{figgs}).  Note that (\ref{clas2d}) differs from conventional $\mathbb{Z}_2$ gauge theory in that the Ising variables are defined on sites rather than bonds.  There is a one-dimensional gauge group~\cite{cenke}, and the ground states are gauge-equivalent to the uniform state.  This ground-state degeneracy may seem topological because the 2D model has a 1D entropy, which scales with the boundary of the model.  However, the ground-state degeneracy is relatively insensitive to the topology of the 2D manifold, unlike the topological degeneracies in e.g. fractional quantum Hall systems.

We instead use the term ``dimensional reduction''.  This term was first used to describe situations where the free energy of one problem in dimension $d$ is exactly equivalent to the free energy of another problem in dimension $d+2$, as in the equivalence between the hard-sphere gas in dimension $d$ and branched polymers in dimension $D=(d+2)$~\cite{imbrie,cardydimred}.  For the 2D classical model (\ref{clas2d}), the free energy per site is the same as that of a 1D model, and there is a similar relationship of correlations.  In the more exotic cases discussed in section II, there are exact properties like duality relations and transition locations that are equivalent for one model in dimension $d$ to another model in dimension $d+n$, although the free energies are demonstrably inequivalent.

The ordering that exists at $T=0$ in this classical model can be viewed as ``bond order'', since all horizontal bonds in the same column are either ferromagnetic or antiferromagnetic, as for all vertical bonds in the same row.  Alternately, since every closed loop on the square lattice has an even number of bonds from each column and row, the order can be defined in terms of the Wilson loop (the product of $s_i$ variables around a closed loop).  Dimensional reduction occurs not just for thermodynamics but also for correlation functions: two vertical bonds on the same row have correlation function
\beq
\langle b_1 b_2 \rangle = (\tanh K)^{|x_1-x_2|}
\eeq
just as for the 1D Ising model.  Two vertical bonds on different rows have zero correlation by symmetry.

\begin{figure}
\includegraphics[width=1.5in]{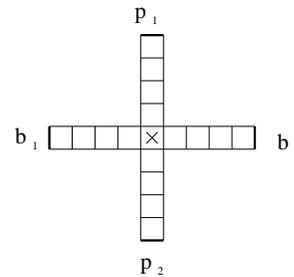}
\caption{The cross-like 4-bond correlation function. The leading
nontrivial term is the product of all the plaquettes above except
the center plaquette (marked).} \label{correlation}
\end{figure}

However, the four-point correlations do not factorize as one might
expect.  The correlation function of two horizontal bonds on the
same column and two vertical bonds on the same row is
(Fig.~\ref{correlation}) \beqn \langle b_1 b_2 p_3 p_4 \rangle &=&
(\tanh K)^{|x_1-x_2| + |y_3-y_4| - 2} \cr &=& (\tanh K)^{-2}
\langle b_1 b_2 \rangle \langle p_3 p_4 \rangle. \eeqn This lack
of factorization indicates that the model does not simply decouple
into a set of independent lower-dimensional systems.  Experiments
are typically most sensitive to the two-point correlations, which
show a strong anisotropy that will be the most dramatic
experimental signature of dimensional reduction.  Usually lattice
systems that respect lattice symmetry are spatially isotropic
sufficiently close to a critical point, but this does not hold
for the examples we discuss.

Our focus in this paper is on physical problems that naturally generate plaquette interactions on the square lattice without ordinary two-point interactions.  However, there are materials of considerable current interest, like the high-temperature parent material La$_2$CuO$_4$, that combine four-spin and two-spin interactions on the square lattice.  Neutron scattering measurements on this material~\cite{coldea} indicate that the ring-exchange coupling $J_\square$ is about 40 percent of the two-spin coupling $J$.
While the precise dimensional reduction we study is specific to the absence of two-spin coupling, some indications survive for finite but small two-spin coupling; work in this direction for a bosonic model $U(1)$ case is in~\cite{paramekanti}.

A helpful way to understand the 2D quantum models we study is through their classical 3D analgoues: these are anisotropic combinations of two-point interactions in the imaginary time direction, with gauge-like four-point interactions in the spacelike planes.  Dualities in some of these anisotropic 3D classical models were studied in a series of papers in the 1980s by Savit and collaborators~\cite{savit2,savit3,savitkt}.  The self-duality in  these models can be understood roughly as follows, for the $\mathbb{Z}_2$ case: since the Ising model in 3D is dual to $\mathbb{Z}_2$ gauge theory, it is plausible that a hybrid of Ising and gauge interactions should be dual to itself.  Dimensional reduction in these models is related to the existence of an infinite but not extensive number of conservation laws.  The solvable Kitaev model~\cite{kitaev} is an example of a 2D quantum model with an extensive number of conservation laws.  The existence of an 1D-infinite number of conservation laws is essential to some models of 2D fractionalized phases; the strong dimensional reduction models are ideal examples of this class for which exact nonperturbative results can be derived.


The outline of this paper is as follows.  Section II introduces three physical examples of partial or complete dimensional reduction, and presents numerical results that confirm previous $1/S$ results~\cite{henley,tchernyshyov} for the Heisenberg model on the checkerboard lattice, an example of weak dimensional reduction.  Section III discusses dualities and phase diagrams of 2D quantum or 3D classical models that show strong dimensional reduction.  In section IV we discuss the correlation functions of these theories in more detail, and in particular the sense in which the correlations are characteristic of a reduced effective dimensionality.  Section V summarizes our main conclusions and some experimentally detectable signatures of dimensional reduction in different condensed matter systems.  We also discuss the relationship between these dimensional reduction cases and the fully solvable Kitaev model.

\section{physical realizations}

This section discusses three different physical problems that lead to a reduction of effective dimensionality, either for the ground state dimensionality (``weak'' dimensional reduction) or for both the ground state degeneracy and the quantum disordering phase transition (``strong'' dimensional reduction).  The three problems are: ordering in superconducting arrays of Josephson-coupled $p \pm ip$ superconductors; the Heisenberg antiferromagnet on the checkerboard lattice; and the bosonic ring-exchange liquid introduced by Paramekanti, Balents, and Fisher~\cite{paramekanti}.  The fundamental significance of the first two cases is that these problems generate four-spin interactions on the square lattice with zero two-spin interactions.  The absence of two-spin interactions is essentially a consequence of symmetry in the first two examples, rather than of any fine-tuning.

First we discuss frustrating phases in two-dimensional superconducting arrays made from grains of $p\pm ip$ superconductors like Sr$_2$RuO$_4$: the angular dependence of tunneling into such a grain leads to an effective Ising Hamiltonian with multiple-spin interactions.  Such angular dependence of tunneling provides the most powerful confirmation that the cuprate superconductors have $d$ order~\cite{tsuei,vanharlingen}.  Complex order parameters like $p\pm ip$ and $d \pm id$ give new physics because the phase measured in Josephson tunneling winds continuously as a function of angle, while real order parameters only generate relative phases of $0$ or $\pi$.  This angular-dependent phase causes several unusual collective effects in periodic two-dimensional arrays.

It was shown in~\cite{moorelee} that,
even in the absence of external magnetic field, there is a
effective flux associated with each elementary loop that sometimes frustrates
the Josephson energy. This flux is determined by the Ising-like
time-reversal order parameters in the grains forming that loop.  The
Ising order parameter specifies which of two degenerate states
related by time reversal exists in a particular grain: for
example, $d+id$ ($\sigma=+1$) or $d-id$ ($\sigma=-1$). The
classical Hamiltonian of such a system takes the Higgs gauge
form
\beq H =
-E_J\sum_{<ij>}\cos(\phi_i-\phi_j-A_{ij}(\sigma_i,\sigma_j)).\label{hg}\eeq
In the above $\phi_i$ is the phase of the superconducting order
parameter in the $i$th grain and $\sigma_i$ is the Ising-like time
reversal order parameter in that grain. The gauge field $A_{ij}$ for $p \pm ip$ superconductors
is given by \beq A_{ij}=
\s_i\theta_i-\s_j\theta_j,\label{gauge}\eeq
where $\theta_i$ is the angle between
the crystalline $y$-axis of the $i$th grain and the normal of the
interface from  $i$ to $j$. Hence the Josephson effect couples the discrete
(Ising) variables $\sigma_i$ to the continuous superconducting phase
variables $\phi_i$.

In this paper we will concentrate on this model square-lattice of $p\pm ip$ grains for two reasons.  First, its two-dimensional quantum phase transition is similar to the {\it one-dimensional} quantum Ising transition or two-dimensional classical Ising transition, and shows the same Kramers-Wannier self-duality~\cite{cenke}.  Second, it shows a type of anisotropic ``bond order'' that also occurs in certain frustrated magnets.  First consider the ground states of this problem.  The frustrating phases disappear on the square lattice if, for every unit square, the four spins $s_i = \pm 1$ satisfy
\beq
\frac{\pi}{2} (s_1+s_2+s_3+s_4) = 2 \pi n, \quad n \in \mathbb{Z}
\eeq
which is equivalent to
\beq
s_1 s_2 s_3 s_4 = 1.
\eeq
Hence the ground states of the system are exactly those of the gauge-like theory discussed in the introduction:
\beq \beta E = -K \sum_\square s^\square_1
s^\square_2 s^\square_3 s^\square_4
\eeq
This differs from ordinary $\mathbb{Z}_2$ gauge theory in that the spin variables live on sites, rather than bonds.  One result is that the gauge group is one-dimensional rather than two-dimensional: the total number of ground states is $2^{L_x+L_y}$ for a system of size $L_x \times L_y$.  A typical ground state configuration is depicted in Fig.~\ref{figgs}.  Note that fixing the spins on one horizontal and one vertical line determines all the remaining spins.  Since each face can be chosen independently to have $\prod_{i \in \square} s_i = \pm 1$, the free energy per face (or per site) is $-\beta^{-1} \log Z = -\beta \log(2 \cosh K)$, just as for the 1D Ising model, where all the bonds .

We will focus on adding quantum effects to the simplified model (\ref{clas2d}), even though this model is only exactly equivalent to the superconducting array problem for the ground states (an extended discussion of the classical physics of such superconducting arrays is in~\cite{moorelee}).  The justification for treating the simplified model is as follows: one can imagine integrating out the superconducting phases to obtain an effective action for the Ising spins.  At zero temperature, where the system is in a ground state,
this still gives a nontrivial set of states, essentially on symmetry grounds.  Our treatment in this paper will start from the simpler model (\ref{clas2d}); while the true spin action is presumably more complicated and includes long-ranged interactions between frustrated plaquettes~\cite{moorelee}, it will have the same gauge symmetries as (\ref{clas2d}), so a reasonable expectation is that similar physics will apply even in the much more complicated array model.

The model becomes quantum-mechanical in the presence of a transverse
magnetic field:
\beq H = -K \sum_\square \sigma^z_1 \sigma^z_2
\sigma^z_3 \sigma^z_4 - h \sum_i \sigma_i^x. \label{quant2d}
\eeq
Here again the $K$ interaction is around a plaquette but now the
spin is a quantum spin-half and the $\sigma$ are Pauli matrices.
This model has an infinite number of conserved quantities, as the total
$S_x$ spin on each row and column is fixed modulo 2.~\cite{cenke}  In the superconducting array realization, the magnetic field $h$ corresponds to tunneling between the two order parameters $p\pm
ip$, which will in the limit of strong tunneling induce a single real order parameter $p_x$.  Applying of pressure in Sr$_2$RuO$_4$ is found experimentally to drive the system toward a real $p$ state, but the explanation of this effect is not yet known.
This quantum Hamiltonian (\ref{quant2d}) was discussed previously in~\cite{cenke}.  There is an exact self-duality between the large $K/h$ and small $K/h$ phases, and a critical point at $K=h$.  In this paper, we will derive some new properties of this model and discuss how its physics is related to that of other examples of dimensional reduction.

Another problem that naturally generates multiple-spin interactions without two-spin interactions is the $1/S$ expansion approach for highly frustrated magnets.  Such magnets can have a large degeneracy of classical ground states, i.e., at $S=\infty$, when the local spin is a classical vector.  This degeneracy is split by quantum-mechanical corrections at order $1/S$: for many cases, the effective Hamiltonian to order $1/S$ is known to have multiple-spin interactions between Ising spin variables that parametrize the degenerate manifold of classical ground states.

\begin{figure}
\includegraphics[width=2.5in]{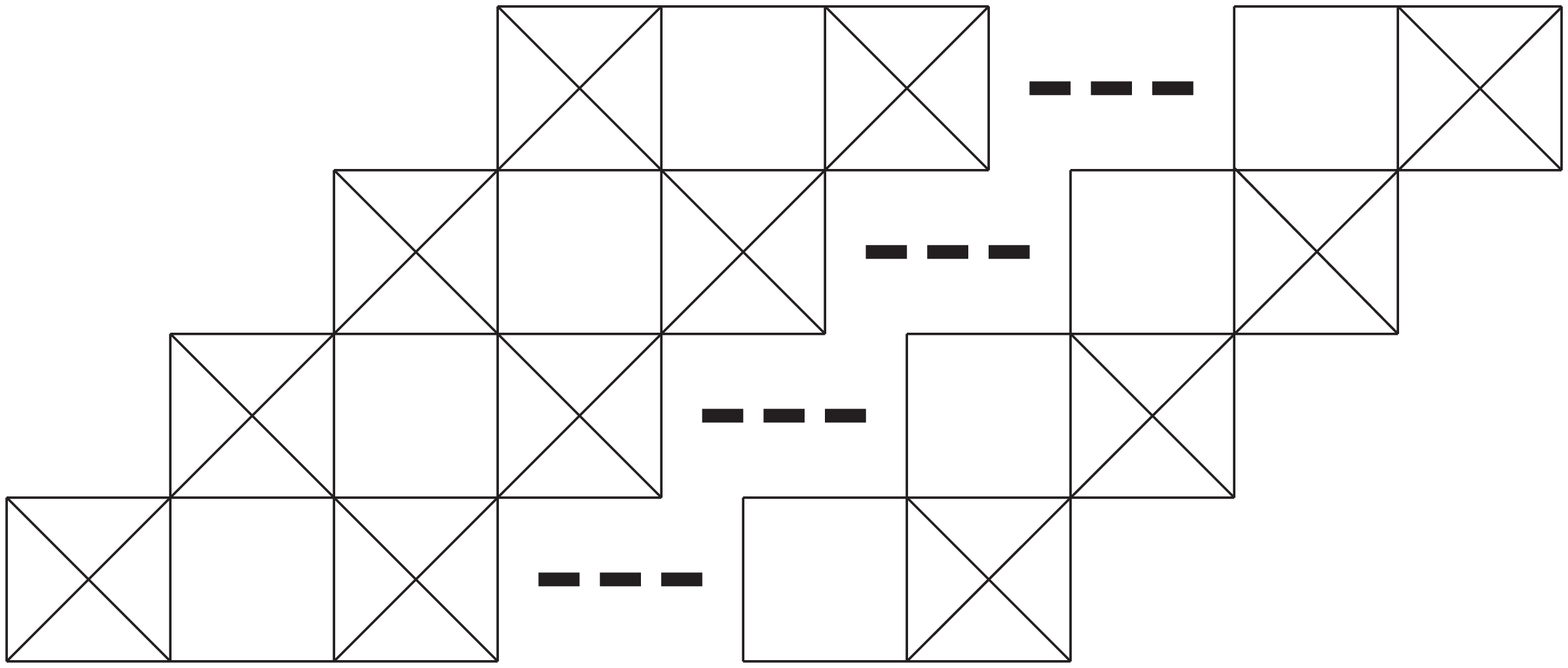}
\caption{Geometry used for the transfer matrix calculations of Table I.  Inclined strips of odd width are terminated by ``tetrahedra'' at both sides.} \label{figone}
\end{figure}

Here we will study one example of this physics in order to understand ``weak'' dimensional reduction.  The 2D checkerboard antiferromagnet (Fig.~\ref{figone}), introduced as a qualitative model for 3D pyrochlore antiferromagnets, has an infinite classical ground state degeneracy.    Although the checkerboard lattice is not known to describe any real 2D system, it is intended to model the corner-sharing geometry of the 3D pyrochlore lattice that describes many important materials.  These ground states can be labeled by Ising variables $s_i = \pm 1$ on a square lattice, subject to a constraint.  The plaquettes with diagonal interactions are like tetrahedra in the pyrochlore lattice, and around any such tetrahedron, the corner spins satisfy
\beq
s_1+s_2+s_3+s_4=0.
\label{conseq}
\eeq

Adding quantum fluctuations at order $1/S$ splits the classical degeneracy, essentially because different classical ground states have different spin-wave frequencies~\cite{henley} (a recent review of this approach is~\cite{tchernyshyov}).  The effective Hamiltonian for these constraints is~\cite{henleyunpub,tchernyshyovmoessner}
\beq
H = - {N J \over S} \sum_\square s_1 s_2 s_3 s_4.
\eeq
This correctly gives the ground states of the checkerboard lattice problem, but is only an approximation for the excited states, since gapless excitations from noncollinear spin configurations are ignored.  We study the Ising spin problem here to understand the role of the constraint (\ref{conseq}), as
the only differences between this problem and the previous model (\ref{clas2d}) is the constraint.  However, the ground-state physics of this checkerboard model is similar: there is a one-dimensional rather than a two-dimensional degeneracy of ground states~\cite{tchernyshyovmoessner}.  Hence there is an effective reduction of dimensionality as far as the ground states are concerned.

We have confirmed numerically that the finite-temperature physics of the two models is very different: even though the ground-state ordering is similar, the checkerboard model does have a finite-temperature phase transition~\cite{tchernyshyovmoessner}, unlike the superconducting array model.  This conclusion is obtained by large-scale transfer-matrix calculations, as unlike the previous case we do not have an analytic solution for finite temperature.  The key to the trivial solution of the previous case was that all the plaquette variables were essentially independent, like the bond variables in the 1D Ising chain; this seems no longer to be the case once the constraint (\ref{conseq}) is added.  The transfer-matrix results on strips of increasing size (Fig.~\ref{figone}) show a peak in the specific heat at $kT \approx 0.95 J$, and this peak sharpens with increasing strip size (Table I).

\begin{table}
\begin{tabular*}{3 in}{@{\extracolsep{\fill}} c   c c c c}
Width&$c_{\rm peak}$&$T_{\rm peak}$&${c^{\prime \prime}(T_{\rm peak})\over c}$&$\Delta T$\\[0.5ex]
\hline\hline\\
9&0.183&0.922&2.62012&0.618\\
11&0.175&0.950&2.78854&0.599\\
13&0.170&0.960&3.05586&0.572\\
15&0.166&0.957&3.32221&0.549\\
17&0.163&0.945&3.48748&0.535\\
\\\hline
\end{tabular*}
\caption{Results of numerical transfer-matrix calculations: estimates for location and width of peak in specific heat for strips of different widths.}
\end{table}

Finally, we briefly review the physical picture introduced in~\cite{paramekanti} that leads to a bosonic ring-exchange model.  The likelihood of strong ring-exchange terms in certain two-dimensional systems has been emphasized in that paper: both spin models and superconducting array models can have many-body interaction terms of the same order as the usual two-body terms.  If the two-body terms can be tuned to zero, as discussed in~\cite{paramekanti}, then the many-body problem can have unexpected physics, including an algebraically correlated phase in 2+1 dimensions.  This is a physical realization of the 3D Kosterlitz-Thouless phase discussed in~\cite{savitkt}.

The Hamiltonian for this model on the square lattice is related to that for ordinary $s$-wave Josephson junction arrays with competition between Josephson and charging energies.  However, instead of a two-grain term $E_J = -J \cos(\phi_i - \phi_j)$, the Josephson coupling is postulated to couple the four spins around a fundamental square.  The resulting quantum Hamiltonian for commensurate filling (an integer number $n_0$ of bosons per site) is
\beq
H = - J \sum_\square \cos(\phi^\square_1 - \phi^\square_2 + \phi^\square_3 - \phi^\square_4) + C \sum_i (n_i-n_0)^2
\label{bosonmodel}
\eeq
Here $\phi_i$ and $n_i$ are conjugate variables forphase and number of the bosons,
and $\phi^\square_i$ refer for $i=1,2,3,4$ to the upper left, upper right, lower right, and lower left grains around a fundamental square.

One point that the reader should keep in mind is that the same ``strong'' dimensional reduction underlies both this bosonic $U(1)$ model and the $\mathbb{Z}_2$ model (\ref{quant2d}).  There is a family of  $\mathbb{Z}_N$ models that interpolates between the two limits, and shows the same two-stage symmetry breaking for large $N$ as the 2D $U(1)$ model with $N$-fold anisotropy.  Some properties of the phase transitions can be determined using the exact duality discussed in the next section.  We find that even though two-point correlations in these theories are highly directional (which allows the physics we find to be measured), these theories do not factor into decoupled 1D quantum problems: rather dimensional reduction is an effective long-wavelength property of some but not all correlation functions.

\section{Dualities and dipole deconfinement transitions in 2D quantum lattice models}

This section starts by explaining how the 2D quantum models with four-site interactions introduced in the previous section show generalizations of standard 2D classical dualities.  These models have the full symmetry of the 2D square lattice.  We show the duality first, for the $\mathbb{Z}_2$ case, directly in the quantum notation, then introduce the associated 3D classical models which were previously shown to be self-dual.  We then derive a picture for the phase transitions of these 3D classical models in terms of a deconfinement of dipoles of solitons, which are dual to the bond variables that order across the transition.   We find that unusual symmetries in the problem lead to a fixed point of lower dimensionality; these symmetries are analogous to the ``sliding symmetry''
that exists in a hydrodynamic model of quantum Hall striped phases.  A recent preprint~\cite{lawler} on this model finds that the sliding symmetry leads to an effectively 1D fixed point.  In this section, we will find a similar conclusion in models that start with the full symmetry of the square lattice.

The self-duality is simplest in the quantum model (\ref{quant2d}).  Just as the duality of the 2D Ising model can also be done in the 1D Ising chain with transverse field \cite{fradkin2}, this quantum duality is the same as that of the classical anisotropic model given below.  First, let us define a dual lattice: $i^\prime$ is on the face surrounded by $i$, $i+\hat{x}$, $i+\hat{y}$ and
$i+\hat{x}+\hat{y}$.  Define new variables on the dual sites by
\beqn S^x_{i^\prime} = \sigma^z_{i} \sigma^z_{i+\hat{x}}
\sigma^z_{i+\hat{y}} \sigma^z_{i+\hat{x}+\hat{y}}.\eeqn Also we
define \beqn S^z_{i^\prime} = \prod_{r\leq i}\sigma^x_r.\eeqn By
$r\leq i$ we mean that both the $x$ and $y$ coordinates of $r$
are not greater than those of $i$.

The new variables $S^x$ and $S^z$ have the same
algebraic relations as $\sigma^x$ and $\sigma^z$, since \beqn
(S^x)^2 = (S^z)^2 = 1,\cr S^xS^z+S^zS^x = 0.\eeqn We can also
check the Hamiltonian in (\ref{quant2d}) can be rewritten in
terms of these new variables as \beqn H = \sum_{i^\prime}-K
S^x_{i^\prime} -
h\sum_{\square}S^z_{i^\prime}S^z_{i^\prime+\hat{x}}
S^z_{i^\prime+\hat{y}}S^z_{i^\prime+\hat{x}+\hat{y}},\eeqn which
has the same form as the original Hamiltonian, except exchanging
$h$ and $K$. This is one example of the quantum version of the dualities.

This quantum version confirms the order that develops at the quantum
phase transition~\cite{cenke} in the $\mathbb{Z}_2$ case.  For example, when $K/h > 1$, the system is in an ordered phase with ``bond order''\cite{cenke}, when $h/K > 1$, the system is in a disordered phase, but it is in the ordered phase in the dual theory. The
order parameter in the dual theory is the dual bond
$S^z_{i^\prime}S^z_{i^\prime+\hat{x}}$, or
$S^z_{i^\prime}S^z_{i^\prime+\hat{y}}$. Written in original
variables, these order parameters can be represented as
$\prod_{r(y)\leq i(y), r(x) = i(x)}\sigma^x_r$, or
$\prod_{r(x)\leq i(x), r(y) = i(y)}\sigma^x_r$. $r(x)$ and $r(y)$
are the $x$ and $y$ components of coordinate $r$. Non-vanishing
expectation value of this order parameter means long range
correlation of $\sigma^x$.

\begin{figure}
\includegraphics[width=2.5in]{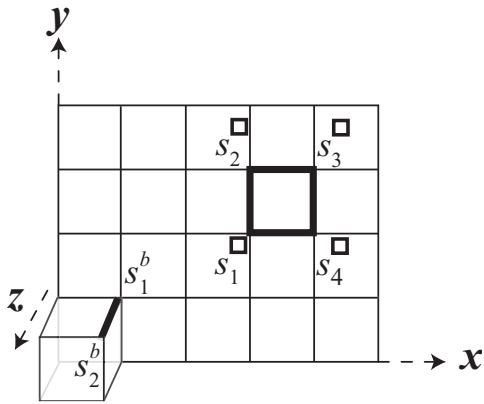}
\caption{The classical anisotropic 3D problem that describes the
quantum critical point of the model (\ref{quant2d}).  The two
types of interactions (shaded bonds) are plaquette interactions in
the planes normal to ${\bf \hat z}$, and bond interactions along
${\bf \hat z}$. } \label{figanisotropic}
\end{figure}

We now introduce a set of classical models in order to find a
field theory description and phase diagram.  The relation between
quantum and classical models for the above $\mathbb{Z}_2$ case is
given in detail in~\cite{cenke}; here we will simply introduce the
classical 3D $\mathbb{Z}_N$ Villain models that were previously
shown to be self-dual~\cite{savit3}.  A slightly different
self-dual $N=2$ model (Fig.~\ref{figanisotropic}) was defined
in~\cite{cenke} to make the notation consistent with the
Kramers-Wannier duality of the 2D Ising model: for Ising variables
$s_i$ on the cubic lattice, \beq Z = \sum_{\{s\}}\exp(K_{xy}
\sum_{\square}s_{i} s_{i+\hat{x}} s_{i+\hat{y}}
s_{i+\hat{x}+\hat{y}} + K_z \sum_b s_i s_{i+\hat{z}}).\eeq the
duality relation $\sinh(2 K_{xy}) \sinh(2 K_z) = 1$ is exactly the
same, but a high-temperature expansion shows that the free energy
is not.  These classical models have an asymmetry between one
direction $t$ and the other two $(x,y)$: define the $\mathbb{Z}_N$
Villain model for a cubic lattice $(x,y,t)$ as \beqn
Z=\sum_{\{q=-\infty\}}^{\infty}\sum_{\{j_\mu=-\infty\}}^{+\infty}
\exp[-\frac{\beta}{2}\sum_i(\frac{2\pi}{N}\triangle_{t}q_i-2\pi
j_{t;i})^2\cr-\frac{\beta}{2}\sum_i(\frac{2\pi}{N}
\triangle_{xy}q_i-2\pi j_{xy;i})^2 ]. \eeqn Here
$\triangle_{xy}q_i$ means \beq
q_{i+\hat{x}+\hat{y}}-q_{i+\hat{x}}-q_{i+\hat{y}}+q_{i}\eeq and
the integer $j_{xy;i}$ is defined on the plaquette surrounded by
$q_{i+\hat{x}+\hat{y}}$, $q_{i+\hat{x}}$, $q_{i+\hat{y}}$,
$q_{i}$.

The coupling in the dual theory is related to that in the original theory by
\beq \beta_{dual}\beta = \frac{N^2}{(2\pi)^2}.\eeq For the Villain model defined
above, the dual theory has exactly the same form as the original theory; demonstrating this self-duality is a simple if lengthy calculation~\cite{savit3}.   If the theory
has two critical points, they are connected by the relations \beq
\beta^c_1\beta^c_2 = N^2/(2\pi)^2, \eeq which is used in the
following several sections.  If it has one critical point, it must lie at the
self-dual coupling $\beta_c = N / (2 \pi).$
For higher dimensions~\cite{cenke}, we can just replace $\triangle_{xy}$ by
$\triangle_{xyz}$, the alternating sum of the eight corners of a fundamental cube,
and most of the following can be generalized to higher dimensions.

The anisotropic $\mathbb{Z}_N$ Potts model, which is a more direct generalization of
our $\mathbb{Z}_2$ model~\cite{cenke}, has similar duality relations. The $\mathbb{Z}_N$
Potts model is \beq
Z=\sum_{q=0}^{N-1}\exp[\beta\sum\cos(\frac{2\pi}{N}\triangle_t
q_{i})+\beta\sum\cos(\frac{2\pi}{N}\triangle_{xy}
q_{i})].\label{zn}\eeq And following the formulation in~\cite{savit} of the self-duality of the 2D Potts model, one finds that for $N = 2,3,4$, the models are self-dual; but for $N \geq 5$, this series of models are not exactly self-dual.

Now we turn to models with $U(1)$ symmetry. In the pure cosine
form (\ref{bosonmodel}), there is not an exact self-duality. The
reason is that the self-duality of $\mathbb{Z}_N$ model is a
duality between the vertex operator and the soliton operator; the
vertex operator is contained in the $U(1)$ model by adding to it
some additional terms, which will make the $U(1)$ model reduce to
a limit of $\mathbb{Z}_N$ ones. And in a pure cosine U(1) model,
there is no vertex operator in the Lagrangian, so it lacks a
natural field that is dual to the solitons. The partition function
of the $U(1)$ Villain model is \beqn
Z=\int_{-\infty}^{\infty}D\theta\sum_{\{j_\mu=-\infty\}}^{+\infty}
\exp[-\frac{\beta}{2}\sum_i(\triangle_{t}\theta_i-2\pi
j_{t;i})^2\cr-\frac{\beta}{2}\sum_i(\triangle_{xy}\theta_i-2\pi
j_{xy;i})^2 ], \label{u1}\eeqn This anisotropic 3D model has a
Kosterlitz-Thouless-like transition~\cite{savitkt,paramekanti} and the
duality relation $\beta_1 \beta_2 = 1$.
The rest of this section discusses this phase transition as a
confinement-deconfinement transition of dipoles of solitons and
obtains a phase diagram for the $\mathbb{Z}_N$ models.

Solitons (a.k.a. monopoles) and other topological defects have behavior that is quite sensitive to dimensionality.  Since the previous section showed that there is a deep connection between quantum ring-exchange models in 2D and classical 2D models, it is natural to ask about how solitons are different from ordinary 2D quantum models.  In the 2D classical XY model, the soliton (vortex) can cause a phase transition, as
can be understood from the famous argument of Kosterlitz and Thouless. A
single vortex of integer winding number $n$ has energy \beqn E_{vortex} =
\frac{\beta}{2}\int^{\pi/a}_{\pi/L} d^2k \frac{n^2}{k^2}\cr = \pi
n^2\beta\ln(L/a).\label{vortexenergy}\eeqn
Here $1/a$  and $1/L$
are ultraviolet and infrared cut-offs.  For simplicity consider the case
with $|n| = 1$. The vortex can take any place within the 2D plane,
so it has roughly $(L/a)^2$ different positions.  The entropy of
a single vortex is thus about $2\ln(L/a)$. So the free energy goes
like \beq F = (\pi \beta-2)\ln(L/a).\eeq
When $\beta < \beta_c = 2/\pi$, the free energy will favor deconfined vortices.

This phase transition does not occur in isotropic 3D models. Because
the energy of a Dirac monopole goes as $\int d^3k 1/k^2$, it is convergent in
the infrared limit, which means the entropy will always dominate; there is no phase
transition and the system is always in a monopole deconfined
phase.
We have seen that certain 3D anisotropic classical models have many properties that
are analogous to a lower-dimensional model. Hence it is not too surprising
that there can be a Kosterlitz-Thouless-like phase transition in the $U(1)$ case: the existence of such a transition was first discussed for anisotropic 3D classical models in~\cite{savitkt}.  Its appearance in 2D quantum models and its physical consequences were studied in~\cite{paramekanti}.   Our goal in the following is to develop a physical interpretation of this transition similar to the vortex-unbinding picture of the 2D Kosterlitz-Thouless transition.  To our knowledge, this discussion in terms of dipoles of solitons is new.  This picture will then be used to obtain the phase diagram of this model with $N$-fold anisotropy.

The energy of a single soliton in the $U(1)$ case, calculated using the Gaussian part of the $U(1)$ theory (\ref{u1}), is \beqn
(2\pi)^2\frac{\beta}{2}\int\frac{dk_x}{2\pi}\int\frac{dk_y}{2\pi}\int\frac{dk_z}{2\pi}
\frac{1}{k_x^2 k_y^2+k_z^2}\cr =
(2\pi)^2\frac{\beta}{4}\int_{\pi/L}^{\pi/a}\frac{dk_x}{\pi}\int_{\pi/L}^{\pi/a}\frac{dk_y}{\pi}\frac{1}{k_xk_y}
= \beta(\ln\frac{L}{a})^2.\eeqn The $(2\pi)^2$ in front of this
equation is inserted because that the charge of soliton is $2\pi$. The
entropy is modified to $S = 3\ln(L/a)$.  Hence the
free energy is \beq F = \beta(\ln\frac{L}{a})^2 - 3\ln(L/a).\eeq
In the limit of large $L$, the energy always dominates, and we can
never get a deconfined soliton.

However, there is another way to drive a phase transition. Instead of considering the
single soliton, let us consider a dipole of solitons.  By dipole we
mean that $m_i$ takes values $n$ and $-n$ on two sites connected by a bond. For example, $m_i = 1$, and $m_{i+\hat{y}} = -1$. The energy
of a fundamental dipole $n = \pm 1$ is \beqn
(2\pi)^2\frac{\beta}{2}\int\frac{dk_x}{2\pi}\int\frac{dk_y}{2\pi}\int\frac{dk_z}{2\pi}
\frac{k_y^2}{k_x^2k_y^2+k_z^2} \cr=
(2\pi)^2\frac{\beta}{4}\int_{\pi/L}^{\pi/a}\frac{dk_x}{\pi}\int_{\pi/L}^{\pi/a}\frac{dk_y}{\pi}\frac{k_y^2}{k_xk_y}
= (2\pi)^2\frac{\beta}{8}\ln L.\eeqn In the formula above, we have
taken $a = 1$; because every derivative
contains $a$ in front of it, so if $a \neq 1$, we can redefine the
lattice in every direction, $x^\prime = ax$, and the theory should
be the same with the case with $a = 1$. Now we can see that the
energy of dipole is not infrared convergent, it also has a
logarithmic divergence, of the same order of entropy.  This indicates that a Kosterlitz-Thouless-like transition is possible, but there is an important difference in that the energy calculation is convergent in the $y$ direction, and hence cutoff-dependent.

The entropy in this case is no longer $3\ln(L/a)$, because we saw that the
energy is convergent in $y$ direction, so the location of the
dipole in $y$ direction does not matter.  Actually, as the $y$-directed
dipoles are uncorrelated if their $y$ coordinates differ, we can say that their
entropy comes only from their location in the $x$-$z$ plane,
so the entropy is of order $2\ln(L/a)$.  Now we will find that there is a critical
value of $\beta$ beyond which the dipoles are confined to be a
quadrupole; otherwise there is a deconfined dipole phase.  The procedure for obtaining this value is more subtle than in the 2D case.

Dipoles with larger moment have higher energy, so they are more easily confined; this applies both to dipoles with higher charge and to dipoles which have bigger distance between their positive and negative charges. We can see that in the calculation above
we have put in a cut-off of $\pi$ for $k_y$.  However,
this only occurs because this model is so anisotropic that the $y$-directed dipole only has correlations in $x$ and $z$ directions: two $y$-directed dipoles can only be correlated when they are at the same $x$-$z$ plane.


With this in mind, we want to find a effective field
theory for (\ref{u1}) that can describe this model near the new
Kosterlitz-Thouless critical point. In the 2D case, in order to
describe the vortices, we can replace the vortex coupling in the dual theory
$i 2\pi m\phi$ by $\cos(2\pi
m\phi)$. When we expand the cosine term order by order, we get the vortex gas with charge $2\pi m$. And because the vortex with charge $2\pi$ is the one with lowest scaling dimension, we need only consider the $|m| = 1$ case, because near the
critical point, vortices with higher charges are all irrelevant.
In our 3D model, the single soliton is never relevant; we only want to
find a term that can reproduce the dipole gas.  The most relevant dipoles
consist of two charges $\pm 1$ located on neighboring sites. So we should replace the soliton-field coupling by $\cos(2\pi \partial_x\phi) + \cos(2\pi\partial_y\phi)$.

Hence we consider as a trial theory of this critical point the Lagrangian density for dipoles
\beqn {\cal L} =
\frac{1}{2\beta}\sum(\partial_{t}\phi)^2+\frac{1}{2\beta}
\sum_\square(\partial_x\partial_y\phi)^2\cr+\alpha\cos(2\pi
\partial_x\phi)+\alpha\cos(2\pi \partial_y\phi).\label{field}\eeqn When the two
cosine terms are relevant, the dipoles deconfine, just as the vortices deconfine in the 2D case: just as $\cos (2 \pi \phi)$ generates vortices in the usual 2D case, $\cos(2 \pi \partial_x \phi)$ generates an $x$-directed dipole.  Our approach will be to argue that because of a gauge symmetry in this Lagrangian, the effective theory is not 3D but 2D, and then calculate various properties including the phase diagram from the 2D theory.

Let us see when the cosine terms are relevant for dipoles in the $x$ direction.  We want to compute the correlation function
$\langle\exp(2i\pi\partial_x\phi(\vec{r}_1))\exp(-2i\pi\partial_x\phi(\vec{r}_2))\rangle$.
For simplicity, we assume that $z_1 = z_2$. This is just a simple
Gaussian integral, and easy to evaluate as long as we pay attention to the cancelling
of the infrared cut-off.  This requirement in the 2D case gives rise to the result that the total charge in any order in perturbation theory is 0, so the total charge is conserved to be 0.  One finds that cut-off independence implies that the correlation can be non-zero only if $x_1 =
x_2$: dipoles are uncorrelated along the direction of the dipole moment.

This arises from a 1D gauge symmetry in
the field theory (\ref{field}). Without the cosine terms, the Lagrangian is
invariant under transformations $\phi\rightarrow\phi+f_1(x)+f_2(y)$,
which means there is a particle number conservation along each row and column. If we
identify $\phi$ as a phase, then $\exp(2i\pi\phi)$ is just the
particle creation operator.  So any two-point correlation function like $\langle\exp(2i\pi\phi(\vec{r}_1)) \exp(-2i\pi\phi(\vec{r}_2))\rangle$ is always zero because it breaks the conservation along some
columns or rows; this is another way to see that the single
soliton is always irrelevant. The first nonzero correlation
function is \beqn\langle\exp(2i\pi\phi(x_1,y_1))\exp(-2i\pi\phi(x_2,y_1))\cr\times
\exp(2i\pi\phi(x_2,y_2))\exp(-2i\pi\phi(x_1,y_2))\rangle. \eeqn
This is a four-point correlation function and the conservation
requires that the four points are four vertices of a rectangle. If
we take $x_2 = x_1 + a$, we come back to the case of dipole
correlation function, and the $x$ coordinate of two dipoles must
be the same.

The field theory in (\ref{field}) seems anisotropic, but when calculating the correlation
function of two $x$-directed dipoles (which must have the same value of $y$), one gets isotropic behavior in the $z$ and $y$ directions.  A na\"ive calculation of the correlation function for any $y_1-y_2$ and $z_1-z_2$, as in the 2D case,
gives \beq \sim
(\frac{1}{y^2+v^2z^2})^{\frac{\beta\pi^2}{2}}+....\eeq This seems to predict that the phase transition occurs when the scaling dimension of the dipole operator is 2, at \beq \frac{\beta\pi^2}{2} = 2 \Rightarrow \beta^c_2 =
\frac{4}{\pi^2}.\eeq  However, this calculation is incorrect:  when we calculated
the scaling dimension of the dipoles, we were using the Gaussian
theory, which means we neglected the quadrupole fluctuations on
microscopic scales. In the 2D XY model, one can neglect the dipole
fluctuations because the scaling dimension calculation is insensitive to
microscopic details, but here the cutoff dependence of the final answer indicates that microscopic fluctuations will modify the critical coupling $\beta^c_2$.

We have not been able to calculate all the effects of quadrupole fluctuation directly. Noticing that the correlation functions of dipoles are only nonvanishing in one row or column, depending on the direction of the dipole, and also that the correlation is isotropic in either the $z$-$x$ directions or $z$-$y$ directions, we guess that the
field theory for the dipolar transition is actually the 2D classical $U(1)$ theory.   This conclusion would be precise if {\it all} correlations were confined to these planes, as a field theory is determined by the set of all its correlations; instead there are higher-order correlations that are not purely planar, as in the Introduction, but these seem to be irrelevant at the dipolar transition.

Under this hypothesis, the leading correlations of the anisotropic 3D theory are described by a set of decoupled planar effective Lagrangians for dipoles:
\beqn {\cal L} =
\frac{1}{2\tilde{\beta}}\sum(\partial_{t}\varphi)^2-\frac{1}{2\tilde{\beta}}
\sum(\partial_x\varphi)^2+\alpha\cos(2\pi\varphi).\label{field2}\eeqn
Here $\cos 2 \pi \varphi$ generates $y$-directed dipoles that drive the transition when this term is relevant, and both sums are over one plane normal to $y$.  There is a similar Lagrangian for $x$-directed dipoles, of course. The renormalized coupling $\tilde{\beta}$ determines the planar scaling dimension of the dipole operator as in the 2D case: \beq\triangle_2 = \pi\tilde{\beta}.\eeq  For ${\tilde \beta} > {\tilde \beta}_c = 2/\pi$, there is a Gaussian phase with effective Lagrangian (now for $y$-directed bonds) \beqn {\cal L} =
\frac{\tilde{\beta}}{2}\sum(\partial_{t}\vartheta)^2-\frac{\tilde{\beta}}{2}
\sum(\partial_x\vartheta)^2.\label{fieldgauss}\eeqn


This hypothesis for the field theory immediately suggests a phase diagram for the self-dual $\mathbb{Z}_N$ models introduced above; since the effective field theory for dipoles is a 2D classical theory, the phase diagram is very similar to the classical 2D case.  First recall the effect of $N$-fold anisotropy on the 2D XY model.  The partition function of the 2D $\mathbb{Z}_N$ model
is \beq
Z=\sum_{\{q=0\}}^{N-1}\exp[\beta\sum_{i}\cos(\frac{2\pi}{N}\triangle_\mu
q_{i})].\eeq From now on, we assume that the lattice constant in
all the directions is 1. Following custom, we replace the models with a
simpler model from the same universality class, the $\mathbb{Z}_N$ Villain model:
\beq
Z=\sum_{\{q=0\}}^{N-1}\sum_{\{j_\mu=-\infty\}}^{+\infty}
\exp[-\frac{\beta}{2}\sum_{i}(\frac{2\pi}{N}\triangle_{\mu}q_i-2\pi
j_{\mu;i})^2]. \eeq The $\mathbb{Z}_N$ model is self-dual, with
the duality relation\cite{savit}\beq \beta_1\beta_2 =
\frac{N^2}{4\pi^2}.
\label{dualrelation}\eeq

Renormalization group analysis predicts that, when $N > 4$, this model has two critical
points~\cite{jose}, and the critical $\beta^{c}$ of the two critical
points have the duality relation $\beta^{c}_1\beta^{c}_2 = N^2/(4\pi^2)$.
One of the two critical points, say $\beta^c_1$, corresponds to the change
in relevance of the symmetry breaking term, and the other one
$\beta^c_2$ can be viewed as the Kosterlitz-Thouless phase
transition, with
\beq \beta^c_2 = \frac{2}{\pi}.\eeq  At low temperature, vortices are confined, the chemical potential of vortices flows to 0 under the renormalization group. But at high
enough temperature, $\beta < \beta^c_2$, vortices become relevant.
So the self-duality is also a duality between vertex and vortex
operator. The $(2\pi)^2$ in the denominator of
(\ref{dualrelation}) is due to the fact that the charge of
a vortex is $2\pi$ in these units.

The cases $N \leq 4$ have only one critical point, whose location is fixed
by the duality relation, $\beta^c = N/(2\pi)$. The reason
is that when we calculated the scaling dimensions of cosine terms,
we neglected the contribution of vortices; this is only consistent
when the phase transition caused by the cosine term takes place in a vortex-confined phase, i.e. $\beta^c_1 > \beta^c_2$. This only happens when $N > 4$, when we
have a Gaussian phase. In the case of $N \leq 4$, the temperature of the Kosterlitz-Thouless point is lower than the temperature that make the cosine term irrelevant, i.e. $\beta^c_1 < \beta^c_2$, which means that there is no Gaussian phase.

Now carrying over these results to our problem is very simple: instead of spins we have bonds, and instead of vortices we have dipoles. First, the phase transition due to dipoles take place when the scaling dimension of the dipole operator becomes 2, from the
last formula in last section we know that this happens at
the renormalized coupling \beq\tilde{\beta}^c_2 = \frac{2}{\pi}.\eeq We can
calculate the scaling dimension of the bonds in terms of
the Gaussian theory, and get the result \beq \triangle_1 =
\frac{N^2}{4\pi\tilde{\beta}}.\eeq The phase transition happens
when \beq\tilde{\beta}^c_1 = \frac{N^2}{8\pi}.\eeq If
$\tilde{\beta}^c_1 = \tilde{\beta}^c_2$, we get the critical value
of $N$, which gives us $N_c = 4$, the same result as in the classical 2D case.
This is consistent with the phase diagram obtained~\cite{numerical} by numerical
simulation.

\begin{figure}
\includegraphics[width=2.5in]{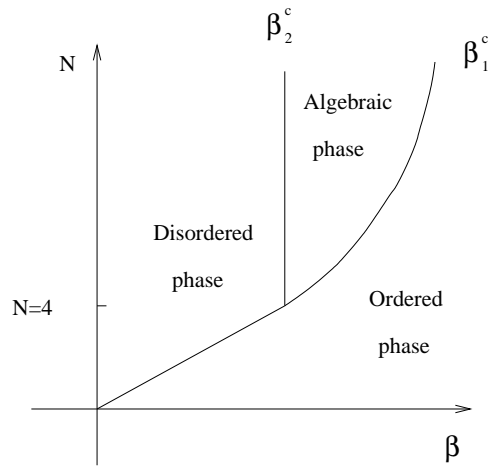}
\caption{The phase diagram of the 3D anisotropic classical $\mathbb{Z}_N$ models:
when $N \leq 4$ there is one critical point, and when $N > 4$ there are two. }
\label{phasediagram}
\end{figure}

The bond-dipole duality is also applicable to another kind of
model. There are numerical results~\cite{savit3} on 3D classical plaquette $\mathbb{Z}_N$ models of the form \beqn H = \sum_i
-K[\cos(\frac{2\pi\triangle_{xy}q_i}{N})+\cos(\frac{2\pi\triangle_{yz}q_i}{N})\cr
+\cos(\frac{2\pi\triangle_{xz}q_i}{N})].\eeqn The bond-dipole
picture is also applicable here: this model also has a
similar gauge symmetry and the ordered phase has a similar "bond
order"~\cite{cenke}, but now with bonds in three different
directions in the classical model, rather than bond order in the $x$ and $y$ directions and ordinary (spin) order in the $z$ direction. We therefore guess that the
effective Lagrangian of bonds is the same as before but with also
$z$-directed dipoles propagating in the $x$-$y$ planes. From this
Lagrangian and its dual form, the critical $N_c$ is still 4,
consistent with numerics~\cite{savit3}.  Hence the phase diagram
(\ref{phasediagram}) is fairly universal for such models.

\section{Correlation functions}

The experimentally accessible physics of the above quantum 2D models is contained in their correlation functions, which exhibit a strong $d$-wave-like anisotropy even near the critical point.  The very simple model presented in the introduction is a good example
of how the correlation functions behave in the disordered phases: the two-point correlations of bonds are localized along rows or columns because of the gauge-like symmetry, but the four-point correlations do not factorize trivially.  In the ordered phases, there is spontaneous symmetry breaking and hence the correlations need not be localized; there are also Gaussian phases intermediate between order and disorder.

Sections I and III have shown that the conservation laws or
symmetries require that the 2-point correlation function between
two vertical bonds, say, should vanish when the 2 bonds are not
along the same row, which was crucial in justifying the
lower-dimensional effective theory. However, the 4-bond
correlation functions are nontrivial even in the disordered phase,
just as in Section I. Taking the $\mathbb{Z}_2$ case as an
example, the classical 3D partition function can be written
as~\cite{cenke}\beq Z = \sum_{\{s\}}\exp(K
\sum_{\square}s_{i}s_{i+\hat{x}}
s_{i+\hat{y}}s_{i+\hat{x}+\hat{y}} + H
\sum_bs_is_{i+\hat{z}}).\eeq First let us calculate the equal-time
2-bond correlation functions $\langle
s_{x_1,y}s_{x_1,y+1}s_{x_2,y}s_{x_2,y+1}\rangle$ in a
high-temperature expansion.  The first term is \beq G_0(x_1,x_2) =
K^{|x_1-x_2|} = e^{-|x_1-x_2|/\xi}.\eeq Here $\xi = -1/(\ln K)$.
This leading term comes from expanding the Hamiltonian along the
shortest path connecting the two bonds. The next leading order
comes from adding one bend to this path, and gives \beq G_1 =
2K^4|x_1-x_2|G_0(x_1-x_2).\eeq

Now let us consider one type of four-bond equal-time correlation: \beq
G^4(x_1,x_2,y,x,y_1,y_2)=\langle b_{x_1,y}b_{x_2,y}p_{x,y_1}p_{x,y_2} \rangle.\eeq Here $p$ denotes a horizontal bond, and $b$ a vertical bond.  Provided that the shortest paths intersect, as in Fig.~\ref{correlation}, the leading term in the high-temperature
expansion is \beq G^4(x_1,x_2,y,x,y_1,y_2) = \frac{1}{K^2}
G_0(x_1,x_2)G_0(y_1,y_2) + \ldots\eeq
So the non-factorization is already present in the leading term. We also checked the
next term:
\beqn G^4(x_1,x_2,y,x,y_1,y_2) =
\frac{1}{K^2} G_0(x_1,x_2)G_0(y_1,y_2)\cr +
10K^4G_0(x_1,x_2)G_0(y_1,y_2).\eeqn Hence the 4-bond correlation
function is not simply $1/K^2\times G(x_1,x_2)G(y_1,y_2)$. The physical picture
in the $\mathbb{Z}_2$ case~\cite{cenke} is that the bond in
the disordered phase is the lowest mobile excitation, but normally the
bond can only move transverse to its direction (i.e., horizontal bonds hop vertically).  However, when two vertical bonds moving in the same row
has opposite momentum, they can interact and scatter into two
horizontal bonds; the horizontal bonds can scatter with each other
and become two vertical bonds. Then in the Hamiltonian there is an
effective interaction between this two kinds of excitations, and
this kind of excitation make the 4-bond correlation function
nontrivial.  Calculating the equal-time correlation of four horizontal bonds, say,
gives the product of two-bond correlations to leading order, but the
next term does not factorize.

The correlations in the Gaussian phase and their connection to conductivity were discussed in~\cite{paramekanti}.   The correlations in the long-range ordered phase are quite different: in this phase, the expectation value of the bond operator is nonzero,
which means the ``gauge'' symmetry is broken.  There is much less
restriction on the physical correlation functions. For example, $\langle
b_{x_1,y_1}b_{x_2,y_2}\rangle$ can be nonzero for different $y_1$
and $y_2$. The physical reason for this is that in the ordered
state is a condensate of the bond variables, and can be regarded as a
reservoir of bonds. So the two vertical bonds in different rows
can interact with each other by exchanging bonds with the
condensate, and the two-bond correlation function is highly
sensitive to the transition between disordered and ordered phases.

\section{Conclusions and other 2D quantum models}

We have shown that frustrated magnets and superconducting arrays naturally generate plaquette interactions on the 2D square lattice, and that such interactions effectively reduce the system's dimensionality, even though the models have the full symmetry of 2D square lattice.  This dimensional reduction can be an exact equality of free energies or an exact equivalence of duality relations, or (``weak'' dimensional reduction) just a ground-state/infinite-coupling property, depending on microscopic details.
The correlations in these theories exhibit the reduced dimensionality, and these theories provide a way to promote phenomena from one quantum dimension, like exact dualities and fractionalized phases, to higher dimensions.

There is a somewhat similar model in two quantum dimensions that is exactly solvable~\cite{kitaev}: for spin-half variables $\sigma$ now defined on {\it bonds} of the square lattice,
\beq
H = -K_1 \sum_\square \sigma^z_1 \sigma^z_2 \sigma^z_3 \sigma^z_4 - K_2 \sum_+ \sigma^x_1 \sigma^x_2 \sigma^x_3 \sigma^x_4.
\label{kitaevmod}
\eeq
The first term contains the four bonds around a plaquette, while the second contains the four bonds meeting at a vertex.  These two terms commute with each other, unlike the two terms in our $\mathbb{Z}_2$ model (\ref{quant2d}), and as a consequence the model (\ref{kitaevmod}) has a 2D infinite set of local conserved quantities and is exactly solvable for all $K_1$ and $K_2$.  The models studied in this paper can be thought of~\cite{cenke} as intermediate between the generic 2D quantum model with no local conservation laws, and the Kitaev model with an extensive set of local conservation laws: ``strong'' dimensional reduction seems to depend on the existence of enough conservation laws that the model's effective dimensionality is reduced, but not to zero.  It is quite remarkable that this exotic behavior occurs in models motivated by experiment.

\section{Acknowledgements}

The authors acknowledge helpful conversations with P. Fendley, E. Fradkin,
D.-H. Lee, A. Paramekanti, and R. Moessner.  J. E. M. was supported by NSF DMR-0238760 and the Hellman Foundation.  Part of this research was carried out at KITP,
supported in part by the NSF under Grant No. PHY99-07949.

\bibliography{../bigbib}
\end{document}